\documentclass[preprint,12pt]{aastex}

\newcommand{\etal}{et al.~}
\newcommand{\bpos}{$B_{pos}$}

\newcommand{\blos}{$B_{los}$}
\newcommand{\jybeam}{Jy beam$^{-1}$}
\newcommand{\mG}{$\mu$G}

\shorttitle{Polarization in Barnard 1}
\shortauthors{Matthews \& Wilson}

\begin{document}

\title{Magnetic Fields in Star-Forming Molecular
Clouds. V. Submillimeter Polarization of the Barnard 1 Dark Cloud}

\author{Brenda C.~Matthews} 
\affil{Department of Astronomy, University of California at Berkeley, 601 Campbell Hall, MC 3411, Berkeley, CA 94720}
\email{bmatthews@astron.berkeley.edu}
\and
\author{Christine D.~Wilson}
\affil{McMaster University, 1280 Main Street West, Hamilton, ON L8S 4M1 Canada}
\email{wilson@physics.mcmaster.ca}

\begin{abstract}
We present 850 \micron\ polarimetry from the James Clerk Maxwell
Telescope toward several dense cores within the dark cloud Barnard 1
in Perseus.  Significant polarized emission is detected from across
the mapped area and is not confined to the locations of bright cores.
This indicates the presence of aligned grains and hence a component of
the magnetic field in the plane of the sky.  Polarization vectors
detected away from bright cores are strongly aligned at a position
angle of $\sim 90^\circ$ (east of north), while vectors associated
with bright cores show alignments of varying orientations.  There is
no direct correlation between the polarization angles measured in
earlier optical polarimetry toward Perseus and the polarized
submillimeter thermal emission.  Depolarization toward high
intensities is exhibited, but toward the brightest core reaches a
threshold beyond which no further decrease in polarization percentage
is measured.  The polarized emission data from the interior envelope
are compared with previously published OH Zeeman data to estimate the
total field strength and orientation under the assumption of a uniform
and non-uniform field component in the region.  These results are
rough estimates only due to the single independent detection of Zeeman
splitting toward Barnard 1.  The uniform field component is thus
calculated to be ${\bf B}_0 = 31$ \mG\ $[\pm (0.52 \hat N - 0.01 \hat
E) - 0.86 \hat z$] in the case where we have assumed the ratio of the
dispersion of the line-of-sight field to the field strength to be 0.2.
\end{abstract}

\keywords{ISM: clouds, magnetic fields, molecules --- polarization
--- ISM: individual (Barnard 1) --- stars: formation --- submillimeter}

\section{Introduction}

Over the past decade, observational evidence for the presence of
magnetic fields in molecular clouds and their role in star formation
has grown dramatically.  However, detection of magnetic fields is not
synonymous with measuring their geometry or the total field strength.
Recent simulations have shown that the low density regimes of
molecular clouds may not be in magnetic equilibrium \citep{pad99},
although in high-density cores, the evidence for magnetic and virial
equilibrium is stronger \citep{mg88,cru99a,basu00}.  While models can help
interpret data, it is still very rare for evidence of Zeeman line
splitting (which traces the line-of-sight field, \blos) and thermal
dust polarization (which traces the plane-of-sky field, \bpos) both to
exist toward regions of similar density in a single cloud.

The Perseus molecular cloud complex is one of the closest star-forming
regions to the Sun.  Its distance is the subject of some debate, but
it is thought to be associated with the Per OB2 association at a
distance of $334 \pm 12$ pc \citep{bb64}.  However, the complex is
likely in front of the OB association \citep{l69}, and \citet{cer85}
suggest that there are in fact two clouds along the line of sight
toward the complex, the second at a distance more comparable to Taurus
at 200 pc.  In this paper, we adopt a distance of 330 pc for the
Perseus association.  CO emission reveals that the complex is
elongated, extending over 55 pc along its major axis (at $\sim
60^\circ$ east of north), but just 15 pc along the minor axis
\citep{sar79}.  Along its length, six denser star-forming clouds
(L1448, L1455, NGC 1333, Barnard 1, IC 348 and Barnard 5) are 
connected by low density molecular gas of $n \sim 10^2$ cm$^{-3}$
\citep{bc86a}.

The Barnard 1 (B1) cloud has been observed in many molecular
transitions and modeled as a multi-phase cloud with a thin outer
envelope, denser inner envelope and a central dense core.
\citet{bach84} observed the cloud in several isotopes of CO, HCO$^+$
and a single transition from NH$_3$ and determined that temperatures
are higher toward the outer edges of the cloud, indicating primarily
external heating.  Three optically visible young stellar objects,
LkH$\alpha$ 327, LkH$\alpha$ 328, and LZK 21 are associated with the
cloud, two of which show IRAS emission.  Three additional IRAS sources
are undetected optically; the presence of these sources indicates some
recent star formation in B1.  In the center of the cloud,
\citet{bach90} observe CS $J=1-0$ emission from dense gas over a
region $\sim 2$ pc $\times$ 5 pc with an accompanying mass of 1200
$M_\odot$. NH$_3$ (1,1) and (2,2) emission toward the CS ``main core''
reveal substantial substructure within the gas, showing evidence for
two or three condensations \citep{bach90}.  IRAS 03301+3057 lies at
the center of the main core, but does not coincide directly with any
of the ammonia peaks; it is located approximately 1\arcmin\ north of
the south-western ammonia peak (the ``southern clump'' of
\citet{bach90}).

High resolution observations of H$^{13}$CO$^+$ by \citet{hir99} reveal
a strong peak (denoted B1-b) at the position of the south-east ammonia
emission detected by \citet{bach90}.  As part of the same study,
continuum emission at 850 \micron\, 350 \micron\ (from the JCMT and
CSO, respectively) and 3 mm (from the Nobeyama Millimeter Array)
clearly identify two high density dust cores within the single
molecular clump.  Based on spectral energy distributions,
\citet{hir99} conclude that these objects are both extremely young
protostars in the Class 0 phase \citep{and93}.  The masses of the
central objects, called B1-bN and B1-bS, are estimated to be no greater
than $7 \times 10^{-2}$ $M_\odot$, indicating extremely young ages of
less than $2\times 10^4$ yr for both sources.  No outflows have been
detected from either source.

One powerful outflow, associated with IRAS 03301+3057 (B1-IRS), has
been identified by \citet{bach90} in CO ($J=2-1$).  It is confined to
a region of 40\arcsec\ (about 0.07 pc).  The dynamical time estimated
from the outflow is $10^3$ to $10^4$ years.  \citet{hir97} measured
the small scale structure of this CO outflow and estimate that the
driving source is very young and is observed in a pole-on
configuration.

The masses of the YSOs seen optically and with IRAS range from 0.2 to
3 $M_\odot$ \citep{bach90}.  Given these low stellar masses, the
stellar to gas mass ratio in B1 is $\sim 0.5$\%, negligibly small even
compared with Taurus where the star formation efficiency has been
recently estimated at 6\% \citep{oni98}.  The observed rotation
velocities within B1 are insufficient to support the cloud against
collapse by a factor of $\sim 8$ \citep{bach90}.  The ages of embedded
but optically visible objects LkH$\alpha$ 327 and LkH$\alpha$ 328 are
between 4-6 $\times 10^6$ yr \citep{ck79}.  Based on this,
\citet{bach90} conclude that a mechanism must be providing substantial
support to the B1 cloud.  The mechanism is generally attributed to a
magnetic field.

Polarization of background starlight from the Perseus cloud due to
selective absorption from dust grains within the complex was measured
by \citet{good90a}, who find that the distribution of polarization
position angles is bimodal, with weaker vectors (less polarized)
aligned along the cloud's projected major axis and stronger vectors
(more polarized) lying roughly perpendicular to the first population.
\citet{good90a} hypothesized that two clouds of differing
magnetic field orientations could be superimposed along the line of
sight.  A prior argument for a second gas cloud along the line of
sight to Perseus and B1 at a distance of 200 pc was presented by
\citet{cer85}.

The B1 cloud has been surveyed for evidence of Zeeman splitting in
dense OH gas more extensively than any other dark cloud.  \citet{lw79}
estimated a 3$\sigma$ limit of 90 \mG\ toward LkH$\alpha$ 327, located
approximately 4\arcmin\ away from B1's strong molecular peak.  B1 was
chosen by \citet{good89} as a strong candidate for magnetic field
detections due to its atypically high non-thermal linewidth
components, and a field strength of $-27 \pm 4$ \mG\ was measured
toward the position of the bright molecular core coincident with IRAS
03301+3057.  (The negative sign indicates that the field is oriented
toward the observer.)  A survey of 12 dark clouds for evidence of
Zeeman splitting yielded only one solid detection -- toward B1 -- with
the 140 ft. Green Bank Telescope \citep{cru93}.  In an observation
toward the source IRAS 03301+3057, a field strength of $-19 \pm 4$
\mG\ was measured, which is consistent with the Arecibo value when
beam dilution is taken into account.

In order to supplement the Zeeman data toward the dense molecular gas
of B1, we have measured polarized emission at 850 \micron\ from dust
toward the ``main core'' of B1 as identified in CS and NH$_3$, both
tracers of high column densities.  Emission from aligned, spinning
dust grains is anisotropic and hence polarized.  Unfortunately,
polarization data reveal no direct information about the field
strength, since the degree of polarization is dependent on other
factors such as grain shape, composition and degree of alignment.  The
degree of polarization is in essence a measure of how effectively the
grains have been ``sped up'' \citep{hil00}.  However, even though the
grain spin is induced by mechanisms other than the magnetic field,
such as the radiation field \citep{dra96} or the production of H$_2$
on the grain surface \citep{pur79}, the magnetic field is expected to
provide the alignment.  Because of this, continuum polarization data
are the principal means of probing the geometry of the magnetic field.
The very sensitive Submillimetre Common-User Bolometer Array (SCUBA)
detector now permits the observation of polarized emission from the
ambient cloud surrounding dense cores.

This paper is the fifth in a series to examine the magnetic field
geometries in star-forming molecular clouds using polarized emission
at 850 \micron.  Barnard 1 is the first dark cloud we have observed,
and these data are the first emission polarimetry toward this region.
The observations and data reduction techniques are described in $\S$
\ref{p5:obs}.  The polarization data are analyzed in $\S$
\ref{p5:poldata}.  We discuss the possible interpretations of these
data and calculate an estimated three dimensional structure for a
uniform field component of B1 in $\S$ \ref{p5:disc}.  Our findings are
summarized in $\S$ \ref{p5:summary}.

\section{Observations and Data Reduction}
\label{p5:obs}

We have used the UK/Japan polarimeter with the SCUBA detector at the
James Clerk Maxwell Telescope\footnote{The JCMT is operated by the
Joint Astronomy Centre on behalf of the Particle Physics and Astronomy
Research Council of the UK, the Netherlands Organization for
Scientific Research, and the National Research Council of Canada.}, to
map polarized thermal emission from dust at 850 \micron\ toward a
dense region of the B1 dark cloud.  The observations were taken from
11 to 13 October 1999.  The polarizer and general reduction techniques
are described in \citet{gre00} and \citet{gre01b}.  To generate a
polarization map, a 16-point jiggle map was made at each of 16
different half-waveplate positions.  After each 12s integration, the
half-waveplate was rotated through $22.5^\circ$ and the mapping
repeated.  The data were flat-fielded, corrected for extinction and
dual-beam corrected using the standard SCUBA software.  Estimation of
systematic errors due to chopping and sky subtraction can be found in
(\citet{mwf01}, hereafter Paper II).  Unfortunately, there are no
large scale 850 \micron\ maps of the B1 cloud in the literature.  This
made the identification of chop angles and sky removal candidate
bolometers more difficult.  The chop angles and throws for each field
center observed are summarized in Table \ref{p5:obs_parameters}.  The
level of precipitable water vapor was very stable over the course of
the observations.  The estimates of $\tau$(225 GHz) from the CSO
ranged from 0.055 to 0.075 over the observations, with 96\% in the
range 0.060 to 0.070.

These data were corrected for an error in the SCUBA clock which placed
incorrect LST times in the data headers during the period from July
1999 to May 2000.  This error did not affect the telescope's
acquisition or tracking, but affects data reduction since the
elevation and sky rotation are calculated from the LST times in the
data headers.  The magnitude of this error over time can be evaluated
and then corrected retroactively as described on the JCMT website.
The error in timing after this adjustment is $\pm 10$s.  The data were
reduced using the Starlink software packages POLPACK and CURSA,
designed specifically to include polarization data obtained with
bolometric arrays.

After extinction correction, noisy bolometers identified for each
night's data were flagged and removed from the data.  Between 3 and 5
bolometers were removed per night.  Prior to sky subtraction, images
were made to examine the flux in each bolometer, since bolometers used
for sky subtraction should not have negative values (produced if one
has chopped onto a location with significant flux, for example).  
The data were sky subtracted using bolometers with mean values close
to zero.  Between 1 and 4 bolometers were used to subtract the sky.
The methods of sky subtraction are discussed in detail in
\citet{jen98}.  Finally, the instrumental polarizations (IPs) were
removed from each bolometer.  All the data sets were then combined to
produce maps of three Stokes' parameters ($I$, $Q$, and $U$), which
were then combined to yield the polarization percentage and
polarization position angle according to the following relations:
\begin{displaymath} 
p = \frac{\sqrt{Q^2 + U^2}}{I}; \ \ \  \theta = \frac{1}{2} \arctan(U/Q).
\end{displaymath}

\noindent The uncertainties in each of these quantities are given by:
\begin{displaymath}
dp= p^{-1}\sqrt{[dQ^2Q^2 + dU^2U^2]}; \ \ \ d\theta = 28.6^\circ / \sigma_p
\end{displaymath}

\noindent where $\sigma_p$ is the signal-to-noise in $p$, or $p/dp$.

A bias exists which tends to increase the $p$ value, even
when $Q$ and $U$ are consistent with $p=0$, because $p$ is forced to be
positive.  The polarization percentages were debiased according to the
expression:
\begin{displaymath}
p_{db}= \sqrt{p^2-dp^2}.
\end{displaymath}

\noindent Future references to polarization percentage, or $p$, refer
to the debiased value.

Absolute calibration is not part of the standard reduction of
polarization data since the percentage polarization is a relative
quantity.  However, from our Stokes' $I$ map, we can estimate fluxes
by using a reasonable flux conversion factor for 850 \micron\ SCUBA
data.  This quantity is dependent on the chop throw used, and for a
throw of 120\arcsec, the standard flux conversion factor is $219 \pm
21$ \jybeam V$^{-1}$ according to the JCMT website.  However, the flux
conversion factor is at least a factor of two greater for polarization
data due to the presence of the analyzer which retards half the
incoming flux on average.  In practice, the flux conversion factor is
$\sim 2.2$ times the standard value due to imperfect transmission
through the waveplate (J.\ Greaves, 2002, private communication).
Hence, we have scaled our Stokes' $I$ map by 480 \jybeam V$^{-1}$.
The associated uncertainty in this factor introduces an uncertainty of
10\% into the resultant fluxes.

Before filtering the data to select reliable polarization vectors, it
was necessary to estimate the effects of sidelobe polarization in the
position of the main beam.  This is a measure of the minimal
believable polarization, $p_{crit}$, given the potential for sources
in sidelobes to produce artificial polarization signals in the central
region of the map (see \citet{gre01b}).  For our worst case scenario
in B1, the flux contributed at approximately 68\arcsec\ from the map
center is 16 times that at the center.  Examination of polarization
maps of Saturn (which has only a small intrinsic polarization $\sim
0.6$\%, with a minimum of $\sim 0.2$\%) of 13 October 1999 reveal that
the relative mean power 68\arcsec\ from the main beam center is 0.0067
compared to the main beam itself.  The mean polarization percentage at
this position on the SCUBA field is 4.3\%, which is a measure of the
instrumental polarization.  The $p_{crit}$ value is given by:
\begin{displaymath}
p_{crit} \ge 2 \times 4.3\% \times 0.0067 \times 16 
\end{displaymath}

\noindent (see \citet{gre01b}) which gives a minimum threshold
polarization of 0.92\%.  Taking into account the intrinsic
polarization of Saturn, this leaves $\sim 0.3 - 0.7$\% arising from
sidelobe polarization.  We have thus selected vectors for which
polarization percentage, $p > 1$\% as reliable data.  The vectors
selected also have an uncertainty in polarization percentage,
$dp<1$\%, and signal-to-noise in polarization percentage $\sigma_p >
3$.  To minimize the systematic effects arising from the possibility
that we have chopped onto a region of polarized emission, vectors are
selected only if they are coincident with Stokes' $I > 20$\% of the
faintest peak in our map.  As discussed in Paper II, if the reference
position has a flux level $\sim 10$\% that of the source peak and
polarized to the same degree, in the final map the position angle at
20\% the peak would be offset from the correct value by $\le
10^\circ$, while the $p$ value is incorrect by at most a factor of
two.  For the brighter peaks, the effects would be considerably
reduced.

\section{850 \micron\ Polarization Data}
\label{p5:poldata}

\begin{figure}
\begin{center}
\vspace*{16cm}
\includegraphics{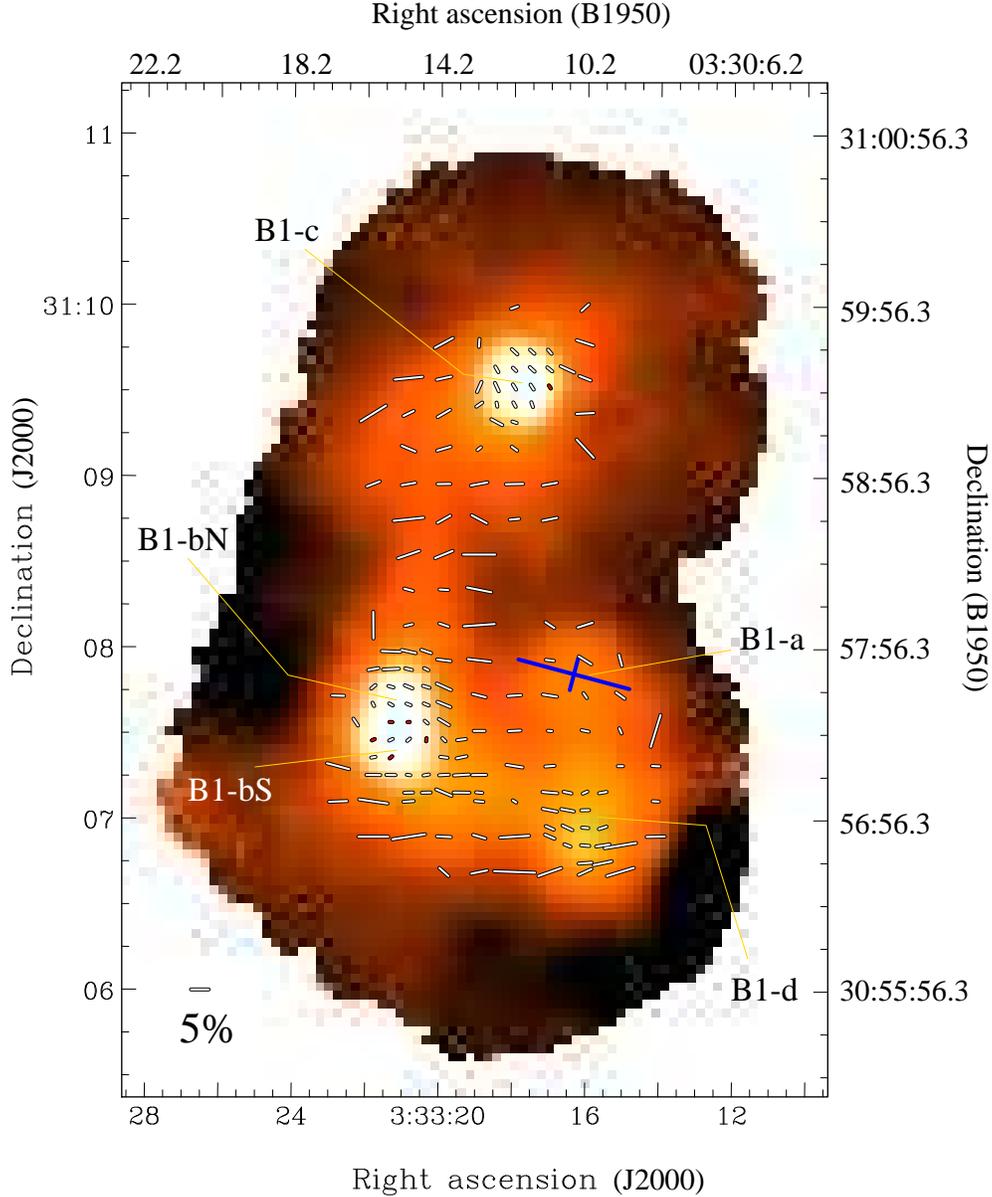}
\caption{850 \micron\ polarization pattern toward the
``main core'' of Barnard 1 is overlaid on the Stokes' $I$ map.  The
greyscale range is $-1\sigma$ to $3\sigma$.  Polarization data were
sampled at 3\arcsec\ and have been binned to 6\arcsec\ (approximately
half the JCMT beamwidth of 14\arcsec) on the bright cores where
signal-to-noise ($\sigma_p$) is high and 12\arcsec\ in the fainter
regions.  All vectors are associated with Stokes' $I$ values greater
than 20\% the B1-d peak flux, $\sigma_p > 3$, and an uncertainty in
polarization percentage, $dp$, $<1$\%.  Red vectors have polarization
percentage, $p$, $<1$\% and were not included in any analysis of the
polarization pattern.  The vectors are accurate in position angle to
better than 10$^\circ$.  The peaks B1-a, B1-bN and B1-bS have been
labeled according to the designations of \citet{hir99}.  We have
designated the other two peaks B1-c and B1-d.  The position of IRAS
03301+3057 is marked in blue where the lines denote the extent of the
uncertainty in the IRAS position.  The mean polarization percentage of
the plotted vectors is 3.6\% with a standard deviation of 2.2\%.
Coordinates in J2000 and B1950 are shown.}
\label{p5:B1map}
\end{center}
\end{figure}

Figure \ref{p5:B1map} illustrates the polarization pattern detected
across the B1 ``main core'' region as identified in CS and NH$_3$ by
\citet{bach90}.  The polarization data are plotted on a colored
greyscale Stokes' $I$ map estimated by summing together the fluxes
detected at all waveplate positions.  Table \ref{p5:allthedata}
contains the data (with $p>1$\%) in tabular form.  Four peaks are
distinguishable, these are labelled B1-a to B1-d.  B1-a and B1-b (N
and S) follow the classification of \citet{hir99} as identified in
H$^{13}$CO$^+$ and 850 \micron\ SCUBA emission.  The presence of two
sources within B1-b was confirmed by 3.0 mm observations with the
Nobeyama interferometer \citep{hir99}.  NH$_3$ (1,1) and (2,2)
emission was observed from peaks corresponding to B1-a, B1-b and B1-c
by \citet{bach90}.  The B1-d 850 \micron\ peak lies approximately
1\arcmin\ south of the B1-a molecular peak \citep{bach90} which is
likely associated with IRAS 03301+3057 (marked by a blue cross on
Figure \ref{p5:B1map}).  No ammonia emission is concentrated at the
B1-d position \citep{bach90}, although a very low signal-to-noise peak
exists near this position in the H$^{13}$CO$^+$ map of \citet{hir99}.

\subsection{Polarization Position Angles}
\label{p5:padata}

\begin{figure}
\begin{center}
\vspace*{14cm}
\includegraphics{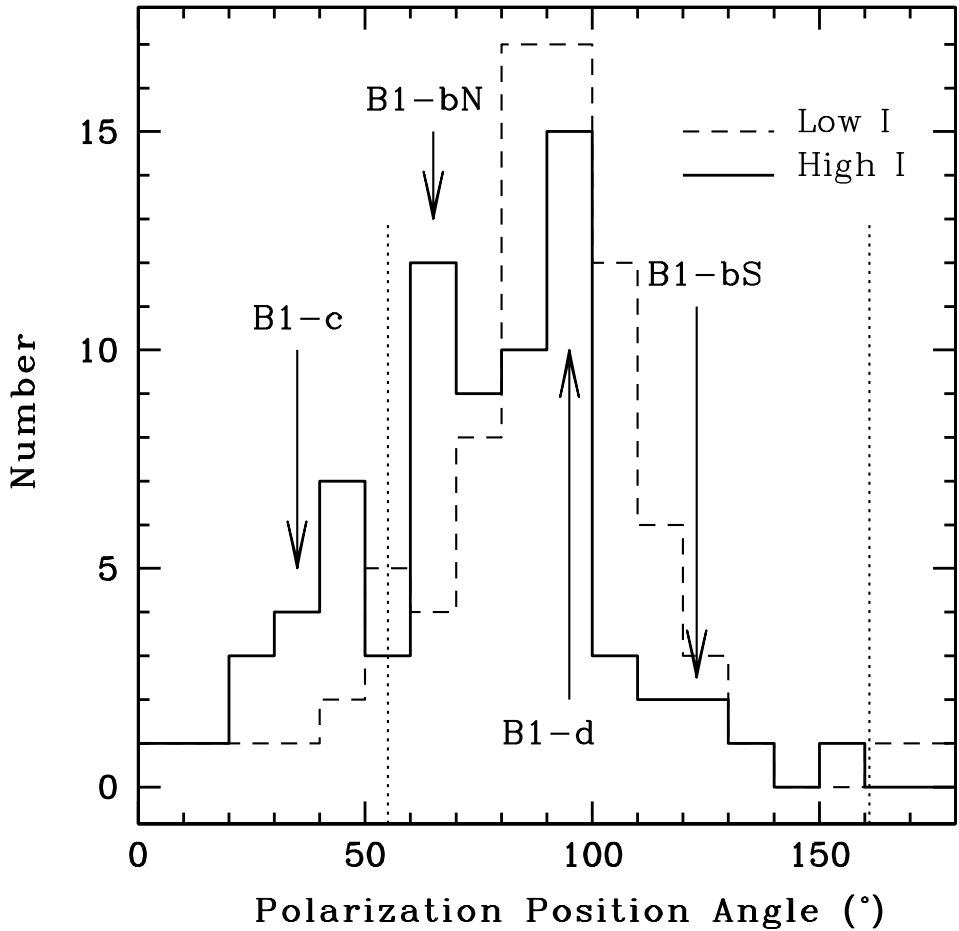}
\caption{Distributions of position angle in B1 are shown
using histograms of the $\theta$ measured toward faint (dashed) and
bright (solid) regions separately.  The threshold used to discriminate
between faint and bright regions was a Stokes' $I$ value of 720 mJy
beam$^{-1}$, as discussed in \S \ref{p5:continuum}.  The vectors
associated with lower column densities exhibit a Gaussian distribution
with a mean of $91^\circ$ and dispersion of $19^\circ$.  The
distribution of vectors toward the cores is not Gaussian, with each
core dominating a different part of the total histogram.  The position
angles are labeled with the core dominating each peak.  The dotted
lines show the emission polarization position angles which correspond
to the peaks observed in the \citet{good90a} distribution of optical
absorption polarimetry (i.e., they are rotated by 90$^\circ$).}
\label{p5:pahist}
\end{center}
\end{figure}

Polarized emission is detected both on the bright cores and in regions
of lower column density between them.  The degree of alignment across
the region is evidence for the presence of ordered magnetic fields
within the main core of the B1 cloud.  The data in high intensity regions
have been binned to 6\arcsec\ sampling, while data in fainter regions
are binned to 12\arcsec\ to improve the signal-to-noise ratio.  The
distributions of polarizations associated with faint emission and
bright emission are plotted separately on Figure \ref{p5:pahist}.

The distribution for faint $I$ (dashed line) is approximately
Gaussian.  A fit to these data yields a mean of 91.3\degr\ with a
distribution width of 19.0\degr.  Polarization position angles are
measured such that values increase east of north.  A goodness of fit
measure to the data yields $\chi^2_{red} = 0.6$.  The statistical mean
of the distribution is 88.3\degr\ (east of north) with a standard
deviation of 27.7\degr.  The distribution of vectors in regions of
bright emission, however, cannot be fit effectively by a Gaussian (or
even a series of Gaussians).  The solid line of Figure \ref{p5:pahist}
shows several peaks, each of which corresponds roughly to one of the
bright peaks.  We have indicated the peak sampled on the distribution.

Therefore, the polarization pattern in the ambient cloud material
around the cores is defined by a mean polarization direction where the
vectors are distributed about 90\degr\ (east of north), while the
cores each exhibit different mean position angles.  The core B1-b
shows systematic variation in position angle.  The northern core has
$<\theta> \sim 65$\degr, while the southern core exhibits $<\theta>
\sim 120$\degr.  The B1-c core has $<\theta> \sim 35$\degr, and the
B1-d core peaks around 90\degr\ east of north, in alignment with the
fainter material in the cloud. This could indicate that this core is
not strongly differentiated from the ambient cloud yet.

On Figure \ref{p5:pahist}, we have also plotted the position angles
associated with the two peaks observed in the optical absorption
polarization data of \citet{good90a}, taking into account the
90$^\circ$ offset expected for emission polarimetry.  These positions
($55^\circ$ and 161$^\circ$) are not coincident with any of the peaks
in the 850 \micron\ distribution, either at high or low intensity.

\subsubsection{Correlations between Adjacent Vectors}

\begin{figure}
\begin{center}
\vspace*{15cm}
\includegraphics{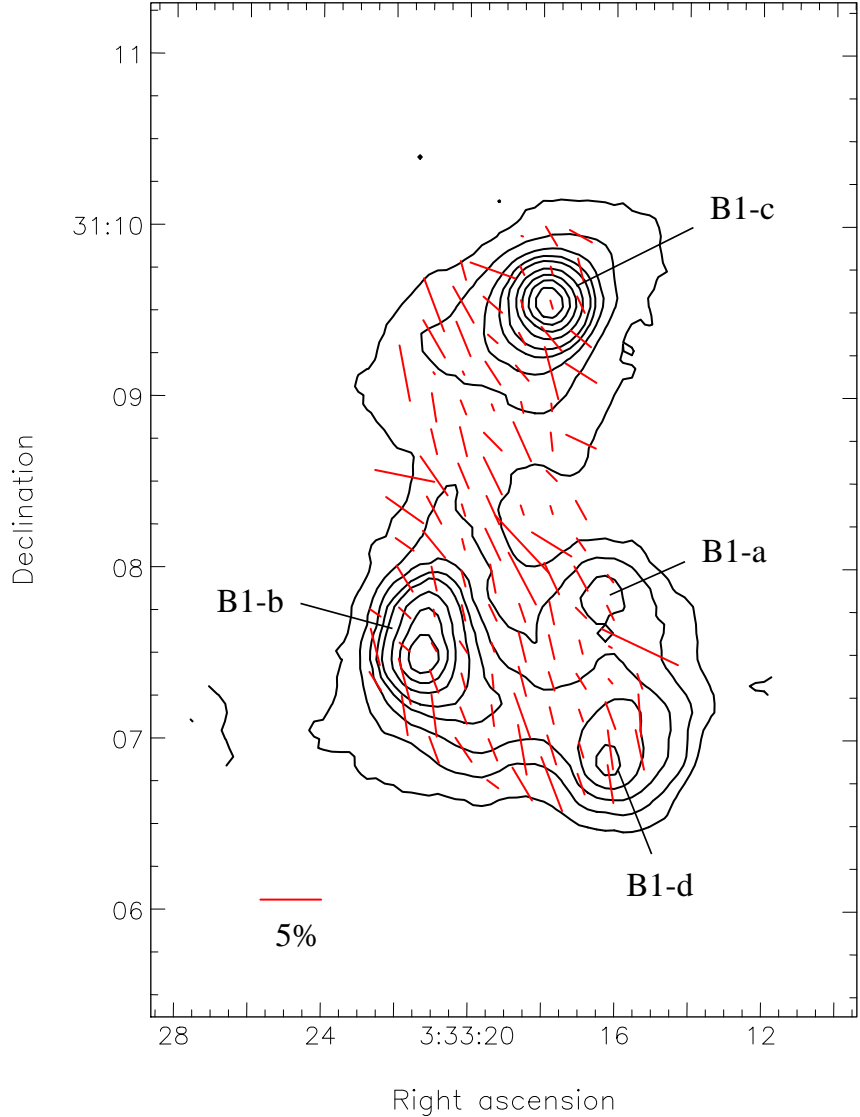}
\caption{Core positions and the variation between
neighbouring vectors.  The positions of the cores are shown with
contours of the Stokes' $I$ emission.  Contours plotted are: 0.2 to
1.0 \jybeam\ in steps of 0.2 \jybeam, and 1.4 to 2.6 \jybeam\ in steps
of 0.4 \jybeam.  (These correspond to visual extinction magnitude
contours of 20 to 100 in steps of 20 magnitudes, and 140 to 260 in
steps of 40 magnitudes.)  We have compared each polarization vector
shown in Figure \ref{p5:B1map} to its nearest neighbours in eight
directions to a maximum radial separation of 17\arcsec.  The data were
then smoothed to 12\arcsec\ sampling.  The length of the vectors
plotted is the mean change $\Delta p$ in the smoothed grid point,
while the vector orientations indicate the mean change in direction
$\Delta \theta$.  Zero change in angle is indicated by 0$^\circ$ (east
of north).  These data reveal that within cores, and between them, the
position angles are strongly consistent, while the large vectors
offsets from $0^\circ$ at the boundaries show where the changes in
vector character occur.  Coordinates are J2000.}
\label{p5:correlate}
\end{center}
\end{figure}

To better examine the changes in the nature of the polarization data
across the map, we have compared each vector to its closest eight
neighbours, calculating the differences in polarization percentages
and position angles for each pair of values.  All the data of Figure
\ref{p5:B1map} were used, except those values for which $p<1$\% (shown
in red).  Next, the results were smoothed onto 12\arcsec\ grid, by
calculating the mean changes in polarization percentage and position
angle.  The resulting map is shown in vector form
in Figure \ref{p5:correlate}, where the vector magnitude is the mean
change in polarization percentage in a grid unit and the vector angle
is the mean change in orientation.  No significant change in
orientation is indicated by a vector at $0^\circ$ east of north.

Total intensity contours illustrate the positions of the cores on
Figure \ref{p5:correlate}.  Only one peak is distinguishable in B1-b,
but it is clear this core is elongated. As expected, the changes
within the cores are relatively small.
The position angles (and even polarization
percentages) are consistent with relatively little change.  Based on
the histogram of the polarization data toward fainter regions shown in
Figure \ref{p5:pahist}, the vectors in this regime were also expected
to be well aligned, and the data of Figure \ref{p5:correlate} show
that this is indeed the case.  The large variations in adjacent grid
units are confined to the edge of the map and the boundaries between
the B1-b and B1-c cores and the fainter material.

\subsection{Continuum Fluxes Toward Cores in B1}
\label{p5:continuum}

Of the four dense cores detected in our polarization map, two have not
been observed previously in continuum emission.  These are B1-c and
B1-d, although the latter may have been confused with B1-a in large
beams (i.e., IRAS) particularly if they are at similar evolutionary
stages.  The brightest source in our map is B1-c, as revealed by the
Stokes' $I$ contours of Figure \ref{p5:correlate}.  We do not detect
both peaks toward B1-b although \citet{hir99} do in an earlier SCUBA
map; their interferometic observation with the Nobeyama Millimeter
Array clearly resolves two peaks toward this source.  The slightly
enhanced emission coincident with IRAS 03301+3057 is the faintest
distinguishable peak at 850 \micron.  Table \ref{p5:B1peaks}
summarizes the peak fluxes toward each of the four cores (for B1-b,
the peak flux is that of B1-bS) and positions of these peaks.

For a cloud at a distance of 330 pc and assuming $\kappa_{850\micron}
= 0.01$ cm$^2$ g$^{-1}$, a mean molecular weight of 2.33 and a dust
temperature of 20 K, we find that the column density $N(H_2) = 10^{23}
S_{850}$[\jybeam] cm$^{-2}$.  The B1-c peak has a flux of 3 \jybeam,
which implies $N(H_2) = 3 \times 10^{23}$ cm$^{-2}$, or 300 magnitudes
of visual extinction.  Toward B1-bS, we estimate $N(H_2) = 2.5 \times
10^{23}$ cm$^{-2}$, which is within 50\% of the estimate of
\citet{hir99}.  Assuming a core depth comparable to the FWHM of
30\arcsec\ for B1-c yields a volume density of $2 \times 10^6$
cm$^{-3}$ which is typical of prestellar and protostellar core
densities.

\subsection{Depolarization in Barnard 1}

Figure \ref{p5:correlate} suggests that changes in polarization
percentage are small within the cores of B1.  The statistical means of
the low intensity and high intensity vector populations are 4.5\% and
2.6\% in 81 and 74 values respectively.  The standard deviations in
these populations are 2.3\% and 1.4\%.  In this case, the
depolarization effect, declining polarization percentage with
increasing intensity, may be weak within parts of B1.  The easiest way
of examining the depolarization effect is to plot $p$ versus $I$ for
all vectors on the polarization map.

In Figure \ref{p5:depol}, we plot the data for the B1 region as
presented in Figure \ref{p5:B1map} excluding only those data values
with $p<1$\% (plotted in red).  The data exist in two populations,
where the data at low intensities are binned to 12\arcsec\ (shown as
crosses) and the data at high intensities are binned to 6\arcsec\
(shown as circles).  Although these plots are shown in log-log space,
the fits to the data were done to profiles of $p$ versus $I$ by
minimizing $\chi^2$.  This is a more effective treatment of the
uncertainties since those for low values of $p$ are exaggerated in log
space.  The fits to these two populations produce completely 
consistent slopes, indicating that both can be characterized by power
laws of the form: $p = A I^\gamma$ with an index of $\gamma \approx
-0.8$.  At high values of $I$ however, there is a slight thresholding
of polarization percentage.

\begin{figure}
\begin{center}
\vspace*{13cm}
\includegraphics{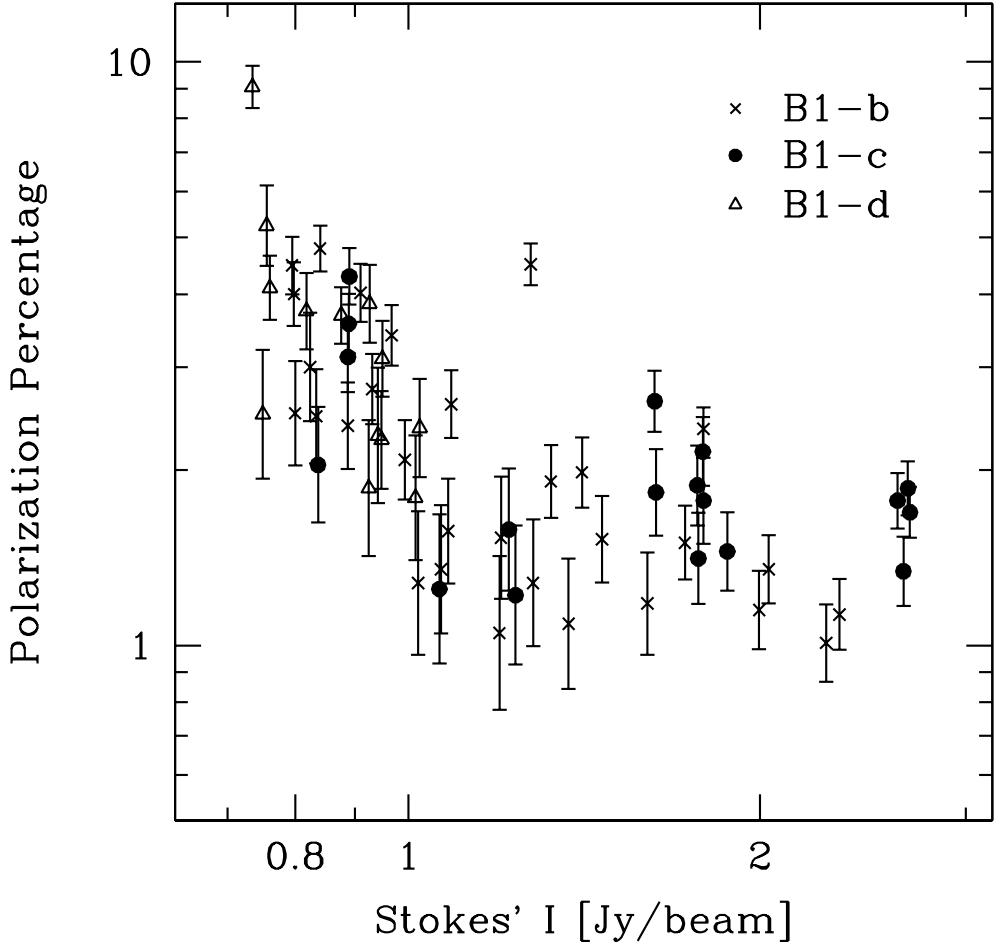}
\caption{Depolarization toward individual cores.  To better
examine the change in polarization percentage at high intensities, we
plot log $p$ versus log $I$ for individual cores.  Essentially, we
take all vectors around each core above the intensity threshold
discussed in Figure \ref{p5:B1map}.  By plotting the three bright
cores separately, it is clear that B1-b and B1-c show higher values of
$p$ than expected given the slope of the relation below 1 \jybeam.
The threshold corresponds to $\sim 30$\% of the peak of B1-c and $\sim
40$\% of B1-b.  In the text, we discuss possible systematic effects
which could explain this flattening (i.e., the truncation of the data
set below 1\% and optical depth). We conclude that the threshold is a
real effect for the B1-c core.}
\label{p5:pvsIcores}
\end{center}
\end{figure}

To better illustrate this, in Figure \ref{p5:pvsIcores} we show the
same style plot for three cores: B1-b (north and south combined), B1-c
and B1-d.  Vectors toward the cores B1-b and B1-c exhibit higher
values of $p$ than expected given the declining trend below 1 \jybeam.
In fact, the distribution of $p$ versus $I$ flattens at high
intensities.  In the case of B1-b, this flattening could be the result
of our lack of sensitivity to values of $p<1$\%.  However, for the
B1-c core, this constraint removes only a single vector (which has a
value of $0.99 \pm 0.25$\%).  Thus, in B1-c, the depolarization effect
does not follow the usual trend of declining polarization percentage
as intensity increases.  (The B1-d core does exhibit depolarization to
its peak, which is significantly lower in intensity than B1-b and
B1-c.)  The B1-c core is well sampled (with only one vector missing)
and definitely exhibits a flat dependence of $p$ on $I$ down to 30\%
of that core's peak.  The omitted vector (shown in red on Figure
\ref{p5:B1map}) does not correspond to the intensity peak of the core,
but is associated with an intensity just 2/3 of the peak observed. To
our knowledge, this is the first case of a bright core which does not
exhibit depolarization over its whole observed intensity range.
The observed threshold is not the effect of optical depth.  We
estimate that the optical depth at the B1-c peak is $\sim 0.015$ which
is $<<1$.  

\section{Discussion}
\label{p5:disc}

\subsection{Interpreting the Polarization Pattern}

Optical polarimetry using absorption of light from background stars
was used to probe the magnetic field structure through dust at low
extinctions in the Perseus cloud complex by \citet{good90a}. They
found a bimodal distribution of polarization vector orientations
toward the complex such that the vectors lie roughly parallel and
perpendicular to the major axis of $\sim 60^\circ$.
They fit two Gaussians to their data set with means $71^\circ$ and
$145^\circ$ east of north and 1 $\sigma$ dispersions of $12^\circ$ and
$8^\circ$, respectively.  Since there is no spatial distinction
between the two populations and evidence in observations of molecular
gas that two different clouds could lie along the same line of sight,
they conclude that the two polarization populations are representative
of two distinct clouds at different distances.  The foreground cloud
was predicted to have a low extinction ($A_V < 1$ mag).  Bimodal
distributions in polarization vectors had been previously noted by
studies toward the Perseus clouds NGC 1333 (\citet{ttc80,vss76}) and
Barnard 5 \citep{jos85}.

In the process of generating an evolutionary model for the B1 cloud,
\citet{cru94} adopt a mean plane-of-sky field direction along the
minor axis of the Perseus (and hence B1) cloud which corresponds to
the polarization distribution centered on $145^\circ$ east of north.
In this case, the \citet{good90a} vectors centered on $71^\circ$ east
of north would most likely be associated with the foreground cloud, at
200 pc distance \citep{cer85}.  If the interior of the dense B1 main
core is threaded with the same field geometry as measured on the
periphery of the B1/Perseus cloud, then a field of mean direction $145
\pm 12^\circ$ should produce polarized emission from dust at a
position angle of $55 \pm 12^\circ$.  Figure \ref{p5:pahist}
demonstrates that there is no peak in the polarization data at 850
\micron\ at $55^\circ$, although polarization vectors of B1-c and
B1-bN fall roughly in the range of the optical polarization dispersion
value.  The second distribution measured by \citet{good90a} peaked at
$71^\circ$, which corresponds to an emission polarization angle of
$161^\circ$.  Thus, there is no component of position angle in our
data set which corresponds directly to either population of the
optical polarization data.  Hence, based on our emission polarimetry,
which is believed to arise only in the regions of dense, cold dust, we
cannot conclude which optical polarization direction is more likely
associated with the Perseus complex.

\subsubsection{Complete Depolarization of the Cores?}

The polarization vectors across the B1-d peak are aligned with the
faint emission polarization angles of $90^\circ$, but the brighter
peaks, B1-b and B1-c, both exhibit different position angles.  Given
the previous observations of depolarization in bright peaks and the
polarization plateau across these two cores, it would be tempting to
think that the cores themselves are completely depolarized and that
one might thus measure the polarization toward a different cloud along
the line of sight at their positions.  There are several reasons why
this is unlikely to be the case.

In most well-sampled cores, depolarization is a non-linear function of
intensity, so the polarization percentage rarely reaches zero (e.g.\
\citet{hen01}; \citet{war00}; \citet{gre99}; \citet{gre94};
\citet{min94}), and if it does become unmeasurably small, as it does
at places in B1-b, then it does so at the highest intensities.
Therefore, one might expect to see a varying position angle across
cores that reflects the varying contribution of the core's polarized
intensity to the vector sum with the second cloud, revealing the
second cloud's position angle only toward the peak of the core.
Furthermore, even in the case where the depolarization might
atypically be complete across a core, this would effectively create a
steeper than usual depolarization effect rather than a flat
distribution of $p$ with $I$ as we particularly observe toward B1-c
(see Figure \ref{p5:pvsIcores}).  The reason is that if the polarized
emission arose in a cloud other than B1 and were unassociated with the
B1-c core (which we know to be in B1 through its associated molecular
emission), the increasing contribution to the total intensity from the
core (despite its contribution of zero polarized intensity) should
produce a variation in $p$ with $I$ which varies exactly as the
increase in intensity across the core.  The only way this could be
avoided would be if, as the intensity across the B1-c core increased,
so did the polarized intensity in the second cloud.  These increases
would have to match each other exactly to produce the flat pattern we
see.  We dismiss this scenario as far-fetched.

In B1, we see not just a single core which exhibits a different
position angle orientation than the ambient cloud, but instead there
are two such cores, where B1-b includes systematic variation in
$\theta$ from north to south.  In order to see the optical
polarizations from a foreground and background cloud in absorption,
\citet{good90a} point out that the extinction of the nearer cloud must
be low ($A_V \le 1$ mag) or have significant fluctuations in order to
see through to the background cloud.  If the extinction in the cloud
is on the order of 1 mag, then the cloud is not self-gravitating and
hence is an unlikely source of 850 \micron\ emission (and certainly
850 \micron\ polarized emission, which is at most on the order of 10\%
of the total flux).  This leaves the possibility of a denser cloud
with significant fluctuations in column density across its projected
surface.

The angular distance between the B1-b and B1-c cores is approximately
2 arcminutes.  At a distance of 200 pc, which is the most probable
location of a foreground cloud given previous molecular line
observations, this corresponds to 0.12 pc, or about the scale of a
self-gravitating core in a molecular cloud.  Since we observe
different position angles in the two bright cores (and a change in
position angle within the B1-b core itself), this would suggest that
fields of varying orientations are exhibited in the foreground
fluctuations in the nearby cloud.  This implies that the fluctuations
in the foreground cloud are comparable to or smaller than the scale of
a starless core.  Since the two bright cores are associated with
NH$_3$ emission (and hence known to be within the B1 cloud),
the fluctuations would have to vary on an angular scale similar to the
separation of cores in B1, despite being at a distance nearly a factor
of two smaller.  We conclude that attributing the polarization seen
against the bright cores B1-b and B1-c to a foreground cloud suggests
an unlikely configuration for a nearby cloud not detected in any dense
tracers such as CS and NH$_3$.

However, the flattening of the $p$ versus $I$ relation in B1-c does
suggest that this core is unusual in some way.  Depolarization within
cores is usually attributed to either changing grain or alignment
physics with increasing density or varying (ordered or disordered)
magnetic field geometry. It is possible that no stellar condensation
has formed at the center of B1-c, in which case a field with a
straight geometry may thread this core.  Also, if no outflow is
present, then theory suggests there should be no disk either, which
means the field is unlikely to be tangled on small scales.  Further
study of this source to search for evidence of outflow or a
protostellar condensation could provide some support for the
suggestion that this is a starless core.  However, this would not
completely explain the pattern, since all the starless cores observed
thus far with SCUBA do exhibit depolarization toward their intensity
peaks \citep{war00}.  The cores observed by \citet{war00} are all
significantly closer than B1 (140-170 pc).  The total flux from one of
these cores (L183) is 2.8 Jy at 800 \micron\ \citep{war94}.
Therefore, cores of comparable brightness, both closer and further
(i.e., LBS 23's cores, \citet{mat01}, hereafter Paper III) than B1, do
exhibit depolarization to the highest intensities.  The $p$ versus $I$
relation of B1-c is thus not easily explained purely by differing
resolutions.

\subsubsection{Polarization in Individual Cores}

It is worth noting that the degree of alignment across the B1-a, B1-c,
and B1-d cores is particularly strong.  Figure \ref{p5:correlate}
shows that these three cores are not significantly asymmetric, so the
polarization patterns cannot be said to align with any preferred axes
of the cores.  The elongated core B1-b is the only one which exhibits
systematic variation in polarization postion angle.  Since this core
is composed of two sources, the polarization patterns could be
different within each core and then we observe their vector sum where
the cores overlap.

\begin{figure}
\begin{center}
\vspace*{13cm}
\includegraphics{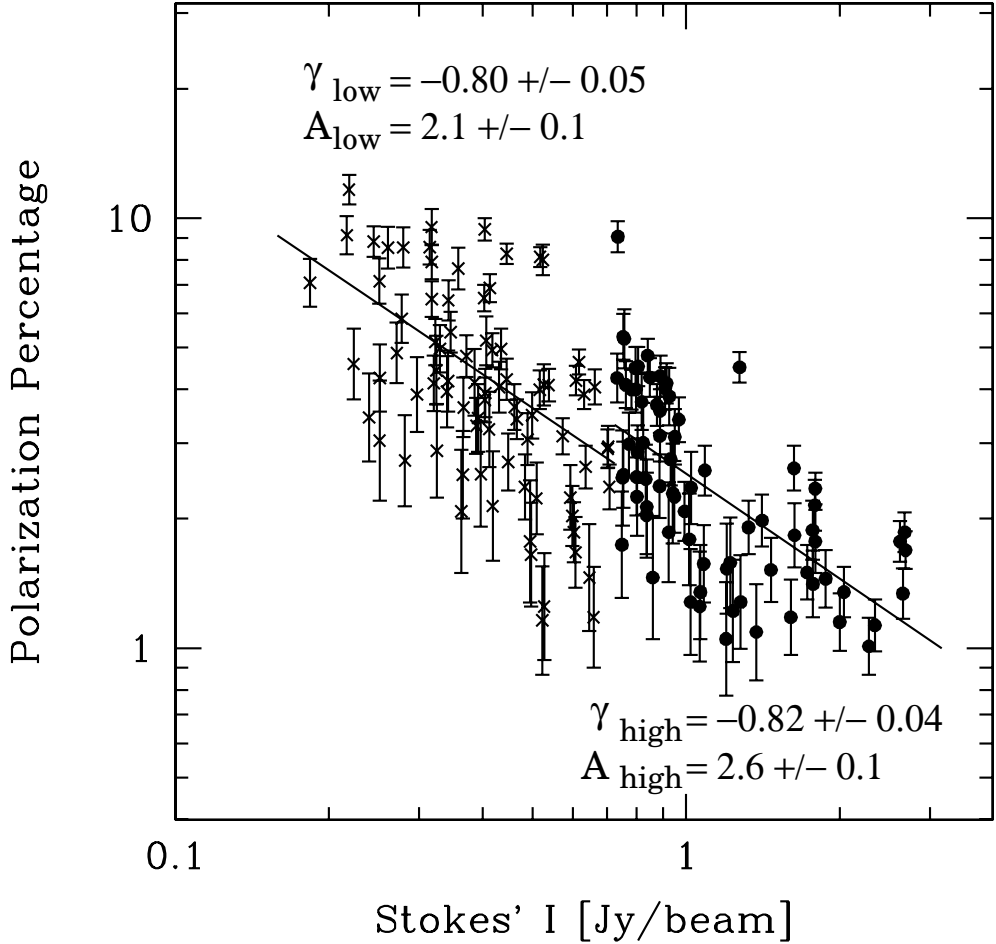}
\caption{Depolarization in regions of low and high
column density.  Polarization is plotted against intensity on a
log-log scale for our two populations of vectors.  Low column density
data are plotted with crosses while circles illustrate
vector magnitudes associated with the cores.  The slopes derived from
fits to the $p$ versus $I$ profiles yield the same power law index,
$\gamma$ (where $p=A I^\gamma$), plotted as the slope on the
logarithmic scale.  We note, however, that despite the scatter in the
plot, the vectors at higher intensities seem to flatten in $p$.  This
can be seen more clearly in Figure \ref{p5:pvsIcores}.}
\label{p5:depol}
\end{center}
\end{figure}

Several models exist which predict the polarization patterns across
cores depending on various physical interpretations.  A recent
publication by \citet{pad01} predicts the continuum polarization for
protostellar cores assembled via supersonic magnetized turbulent flows
in models of molecular clouds.  They find that the universally
observed trend of declining polarization percentage with increasing
intensity in star-forming cores can be reproduced by their model if
grains are aligned only up to a threshold extinction ($A_V \sim 3$ mag
in their simulation).  Several of the resulting $p$ versus $I$ plots
do suggest a flattening in the distribution at high values of
intensity.  However, one difference between their simulations (see
Figs.\ 6 and 7 of \citet{pad01}) and Figure \ref{p5:depol} is the
population of the bottom left quadrant (low $p$ and low $I$); vectors
exist in this region in the simulations, whereas this region is devoid
of vectors in Figure \ref{p5:depol}, despite complete sampling down to
the intensity threshold.  Measurement of low levels of polarized
emission toward lower intensities will become possible with the next
generation of detectors (e.g.\ SCUBA-2).

The fact that there is a potential for confusion between two clouds
along the line of sight makes interpretation of the polarization data
more complicated in Perseus.  However, constraints on the column
density and spatial scale of potential variations in extinction of a
second cloud suggest that the overwhelmingly predominant source of
polarized emission is the B1 cloud and its associated cores.  The
models of \citet{pad01} and \citet{fp00d} relate to cores forming from
lower density, filamentary structures.  B1 is not a significantly
elongated cloud (although the Perseus complex itself appears elongated
on very large scales), and its cores appear to be forming from density
enhancements in its cold interior.  A recent model of magnetized cores
predicts a relation between the geometry of a core and the measured
polarization position angles (e.g.\ \citet{basu00}).  However, the B1
cores are not significantly elongated; the only asymmetry is in B1-b,
which is resolved into two sources with separate extended envelopes by
\citet{hir99}.

\subsection{An Estimate of the Total Field Strength and Direction in B1}

\citet{mg91} describe a method by which the total field strength and
direction can be estimated in a cloud toward which a series of
independent Zeeman measurements of \blos\ and polarization
measurements to infer the orientation of \bpos\ have been made.  Their
formalism is described for absorption polarimetry, but is easily
adapted to emission polarimetry if simple assumptions are made about
the relation between the orientations of polarization vectors and the
plane-of-sky magnetic field.  

A distribution of polarization data which can be fit by a single
Gaussian is a good approximation to the precise function described in
\citet{mg91} in the case where the nonuniform field is relatively
small and the three dimensional random case can be assumed.  As
described in $\S$ \ref{p5:padata} above, the polarization vectors
associated with the faint lower density gas off the B1 cores can be
fit with a single Gaussian distribution which has a mean of 91$^\circ$
(east of north) and a width of 19$^\circ$.  At 12\arcsec\ sampling,
these vectors are not completely independent (since the JCMT beamwidth
is 14\arcsec), but we use them to get a rough estimate of the field
properties.

The Myers \& Goodman model is based on the presence of both a uniform
($B_0$) and non-uniform ($B_n$) component to the field in the
region. Non-uniform in this case refers to a disordered, possibly
turbulent, field component (as opposed to an ordered, but not
unidirectional field). Given a distribution of polarization data with
mean polarization angle, $\Theta$, (in degrees) and dispersion, $s$,
(in radians) and a series of Zeeman measurements with mean uniform
component $\overline B_{0_z}$ and dispersion $\sigma_{B_z}$, the
following quantities can be estimated: the inclination,
\begin{equation}
i = \arctan \left ( \frac{\sigma_{B_z}}{s \overline B_{0_z} } \right ) ;
\label{p5:i}
\end{equation}

\noindent the total uniform mean field strength,
\begin{equation}
B_0 = \frac{\overline B_{0_z}}{\cos{i}};
\label{p5:B0}
\end{equation}

\noindent and the dispersion in $\bf{B}$,
\begin{equation}
\sigma_B = \sigma_{B_z} N^{1/2}
\label{p5:sigB}
\end{equation}

\noindent where $N$ is the number of correlation lengths of the
non-uniform field component through the cloud.  The correlation length
is an expression of how quickly the non-uniform component of the field
changes through the depth of the cloud.  Values of $B_n$ separated by
less than a correlation length are likely to be correlated, while
those more spatially separated than a correlation length are
independent.  \citet{mg91} derive the relation:
\begin{equation}
N_{max} = 69 \left( \frac{A_V}{\rm{mag}} \right)  \left( \frac{B}{\mu \rm{G}}\right)^{-1}.
\end{equation}

\noindent The maximum number of correlation lengths can be estimated
under the assumption that the magnetic and gravitational energy
densities are equal.  Under this condition, $A_V/B \approx 0.17$ mag
\mG$^{-1}$, where $B$ is the total field strength \citep{cf53,mg88}.
This estimate yields $N_{max} \approx 12$.

The r.m.s.\ field strength and the relative strengths of the
non-uniform to uniform magnetic field energy densities can then be
calculated according to the relations:
\begin{equation}
<B^2>^{1/2} = \left ( B_0^2 + 3 \sigma_B^2 \right )^{1/2} ;
\label{p5:Brms}
\end{equation}

\noindent and
\begin{equation}
\frac{M_n}{M_u} = \frac{3 \sigma_B^2}{B_0^2}
\label{p5:ratio}
\end{equation}

\noindent where we have assumed a three-dimensional non-uniform field
component.  

We use the detection of $-27 \pm 4$ \mG\ toward the position of IRAS
03301+3057 \citep{good89} as an estimate of the $B_{0_z}$ value.  In
the absence of other independent detections, there is no estimate of
the dispersion in the line-of-sight field strength.  Thus, we adopt a
parameterized approach to combining the Zeeman and dust polarization
data to estimate the three-dimensional field, where the ratio of
$\sigma_{B_z}/B_{0_z}$ takes on a range of values (0.2, 0.4, 0.6 and
0.8).  Table \ref{p5:Bproperties} then summarizes properties of the
magnetic field in B1 based on equations (\ref{p5:i}), (\ref{p5:B0}),
(\ref{p5:sigB}), (\ref{p5:Brms}), and (\ref{p5:ratio}) for $N=1$ and
$N=10$.  

Finally, the direction of the uniform component of the magnetic field
can be estimated:
\begin{equation}
\hat B_0 = (\hat N \cos \Theta_B + \hat E \sin \Theta_B) \sin i + \cos
i \hat z .
\label{p5:Bdir}
\end{equation}

\noindent For emission polarization data, we assume the mean magnetic
field direction, $\Theta_B$, is related to $\Theta$ by $\pm 90^\circ$.
Emission local to each dust grain should be related to the local field
direction in this manner \citep{hil88}, but for this to be the case in
the vector averaged sum of all polarizations through the cloud (which
is what we measure at the telescope) is to assume that the magnetic
field orientation does not substantially vary through the depth of the
cloud.  This has been shown to be a poor assumption in some regions,
where the polarization data support a more complex field geometry
(i.e., OMC-3 in Orion A, Paper II; NGC 2024 in Orion B, Paper III; and
NGC 2068, Paper IV).  However, B1 is a dark cloud and not part of a
giant molecular cloud complex like Orion; therefore, a uniform field
structure (at least away from the cores) is not an unreasonable
first-order assumption here.  We note however that in utilizing this
relation in B1, we are also assuming that all the polarized emission
is arising in the B1 core, as opposed to in a second cloud as
discussed above.

Using equation (\ref{p5:Bdir}) and $\sigma_{B_z}/B_{0_z} =0.2$ (hence 
$i = 31^\circ$ from the line of sight),
\begin{equation}
{\bf B}_0 = 31 \mu G \left [\pm (0.52 \hat N + 0.01 \hat E) - 0.86 \hat z \right].
\label{p5:B1_Bdir}
\end{equation}

\noindent We note that the direction of the field
along the line of sight is toward the observer.

The field components can also be expressed in terms of two dimensions,
$\hat x$ and $\hat z$, where $\hat x$ lies along the plane-of-sky mean
field direction, as estimated from the mean polarization direction.
In this case,
\begin{equation}
B_{0_x} = B_{0_z} \tan{i}
\label{p5:B0x}
\end{equation}

\noindent and substitution of $B_{0_z}$ and $i$ gives:
\begin{eqnarray}
{\bf B}_0 & = & B_{0_x} \hat x + B_{0_z} \hat z \\
& =  & \pm 19 \ \mu \rm{G} \ \hat x - 27 \ \mu \rm{G} \ \hat z.
\label{p5:B_2d}
\end{eqnarray}

Based on comparisons with theoretical predictions based on the
assumptions of equality of magnetic fields and kinetic energy and
equivalence of magnetic and gravitational energies, \citet{cru93}
found that the $\hat z$-field component measured in B1 agreed well
with predicted values if the inclination of the field to the line of
sight are close to zero.  Furthermore, because B1 was the only cloud
with a detected magnetic field out of 12 in their survey, there was a
concern that B1 might have an atypically strong field.  Based on
statistical analysis, \citet{cru93} concluded that this need not be
the case if the magnetic field in B1 lies nearly along the line of
sight.  The $B_{0_z}$ value of $-27 \pm 4$ \mG\ may be substantially
higher than the magnetic field strength at locations away from the
dense cores.  Therefore, our analysis represents upper limits to the
field strenghths in low column density gas.

However, our detection of ordered polarization vectors from dust
associated with the main core of molecular gas in B1 indicates that at
least some of the magnetic field in the region lies in the plane of
the sky.  Unfortunately, there is no way to unambiguously determine
the plane-of-sky field strength from polarization data alone, since
the degree of polarization may depend on grain size, shape,
composition and degree of alignment or spin, as well as field strength
\citep{hil00}.  We note that the OH Zeeman measurements can
effectively probe regions with densities as high as $n(H_2) \sim 10^4$
cm$^{-3}$ \citep{cru94}, quite comparable to the densities associated
with the dust emission.  Assuming a temperature of 20 K for the main
core, the 850 \micron\ flux density (at a level of 0.2 Jy) implies a
column density of $N(H_2) \sim 2 \times 10^{22}$ cm$^{-2}$, which
corresponds to a density of $n(H_2) \sim 3 \times 10^4$ cm$^{-3}$ if
the emitting dust extends over the main core diameter of 0.8 pc
\citep{bach90}.  In reality, the emitting region of dust may be more
confined along the line of sight, which would imply even higher
densities.  This is certainly the case where flux densities are high.
The column density toward B1-c is $N(H_2) \sim 3 \times 10^{23}$
cm$^{-2}$.  Since Zeeman data have been obtained toward regions of
high density, it is possible that the OH Zeeman data and the dust
emission polarimetry do not arise in precisely the same spatial
regions of the cloud.  They could, therefore, be sampling different
field geometries, or at least different total field strengths.
However, we have applied the \citet{mg91} method over a size scale
comparable to the Arecibo beam at the frequency of the OH Zeeman
measurements of \citet{good89}, making our calculation quite reasonable
for the lower column density dust.

\section{Summary}
\label{p5:summary}

We have detected polarized emission at 850 \micron\ arising from the
dense interior of the Barnard 1 dark cloud.  Our observations are
centered on the molecular ``main core'' observed by \citet{bach90}, in
which three ammonia peaks were identified.  Submillimeter emission is
detected coincident with each of the ammonia peaks.  In total four
dust cores are identified, one of which has been resolved into two
sources \citep{hir99}.  Two of the dust condensations, B1-c and B1-d,
have not been previously observed in continuum.  The B1-a core is
likely the 850 \micron\ counterpart of IRAS 03301+3057.  This source
appears quite faint at long wavelengths.  We note that the detection
of two new dense dust condensations, plus the B1-b binary sources
identified as young Class 0 sources by \citet{hir99}, increases the
number of YSOs or pre-protostellar objects in B1 by almost a factor of
2.  This indicates that the star formation efficiency in B1 is likely
much larger than the 0.5\% estimate by \citet{bach90} since there may
be other as yet unobserved pre-protostellar or protostellar cores in
the cloud of which we have observed only a fraction.

The polarized emission can be separated somewhat arbitrarily into two
sub-groups by the coincident flux levels.  Strong polarizations are
measured toward faint dust emission regions where the mean
polarization percentage is 4.5\% (standard deviation 2.3\%).  The
position angles are distributed about 90$^\circ$ (east of north) and
can be fit by a Gaussian of mean 91.3$^\circ$ and dispersion
$19^\circ$.  The polarizations associated with high intensities (i.e.,
the cores) show smaller polarization percentages, with a mean of 2.6\%
and standard deviation 1.4\%.  The vectors show alignment across the
cores, but each core does not exhibit the same mean position angle.  A
comparison of each vector to adjacent values shows that vectors are
strongly aligned with their neighbours in position angle.  The largest
discrepencies are observed at the ``boundaries'' between the dense
cores and the lower column density dust emission in which they are
embedded.

Over the whole mapped area, we see evidence of the depolarization
observed toward many star-forming cores.  Interestingly, when the
polarization percentages are plotted against intensity for individual
cores, the $p$ versus $I$ relation flattens out at 30\% and 40\% of
the peak emission from the B1-c and B1-b cores, respectively.  The
B1-c core exhibits only one vector with a value of $p<1$\%, and thus
the flattening in that core is real and not just an artifact of a
lower limit on detectable polarizations or increasing optical depth.
The observation of depolarization at the highest intensities of cores
closer \citep{war00} and further (Paper III) than B1 makes it unlikely
that this effect is directly related to our resolution of B1 at 330
pc.

None of the orientations of polarization vectors measured by SCUBA are
directly related to the two mean magnetic field directions detected
with optical polarimetry of the Perseus complex \citep{good90a}.  In
the case where the B1 dense cores could be completely depolarized, the
polarized emission along the line of sight to those cores would arise
completely in the foreground cloud (proposed to be at 200 pc).  If
such a cloud contains fluctuations in extinction, those fluctuations
must be on scales similar to the separation of cores in B1 at
approximately half the distance.  This is required to account for the
differing orientations measured in the two bright cores, B1-b and
B1-c.  We dismiss this scenario as unlikely since, unless the
polarized emission from the foreground cloud rises in such a way as to
offset the increasing intensity toward the B1-c core peak, we should
see a steeper depolarization toward B1-c than in typical cores, not
the threshold we observe.

Finally, following the method of \citet{mg91}, we have estimated the
net field geometry in the B1 main core using our polarized emission
data toward faint regions (centered on $90^\circ$ east of north) and
the line-of-sight field strength toward B1 measured by \citet{good89}.
We find that the total uniform field component is described by:
\begin{equation}
{\bf B}_0 = 31 \mu G \left [\pm (0.52 \hat N - 0.01 \hat E) - 0.86
\hat z \right] , 
\nonumber
\end{equation}

\noindent under the assumption of $\sigma_{B_z}/B_{0_z} =0.2$.  The
ratio of the magnetic energy of the non-uniform component of the field
to the uniform component ranges from 0.09 to 0.9 for this case,
depending on the number of correlation lengths of the non-uniform
component through the cloud.  This result is roughly consistent with
the theoretical predictions based on virial and magnetic equilibrium
in the cloud, for which the line-of-sight field was comparable to the
total predicted field.  The high degree of ordering in the
polarization data itself suggests that some component (possibly a
significant amount) of the magnetic field could lie in the plane of
the sky.

\acknowledgements

The authors would like to thank J.~Greaves, T.~Jenness, and
G.~Moriarty-Schieven at the JCMT for their assistance during observing
and with subsequent data reduction.  A.~Goodman provided clarification
regarding the status of Zeeman observations toward B1, suggestions on
the interpretation of patterns across the cores, and a detailed
referee report which led to significant improvements to this paper.
G.~Petitpas and J.~Wadsley provided helpful discussions on the most
effective means of comparing adjacent vectors.  The research of CDW is
supported through grants from the Natural Sciences and Engineering
Research Council of Canada.  This work was partially supported by NSF
grant AST-99811308.

\begin{deluxetable}{clc}
\tablecolumns{3} 
\tablewidth{0pc} 
\tablecaption{Observing Parameters for Jiggle Mapping of B1} 
\tablehead{ \multicolumn{2}{c}{Pointing Center} &
\colhead{Number of} \\ \colhead{R.A. (J2000)} & \colhead{Dec. (J2000)}
& \colhead{Times Observed} } 
\startdata 
$03^{\rm h}33^{\rm m}$17\fs9 & $+$31\degr09\arcmin32\farcs3 & 16 \\ 
$03^{\rm h}33^{\rm m}$19\fs6 & $+$31\degr08\arcmin28\farcs25 & 12 \\ 
$03^{\rm h}33^{\rm m}$18\fs3 & $+$31\degr07\arcmin03\farcs8 & 30 \\ 
\enddata 
\tablecomments{The chop throw used for all observations was
120\arcsec\ at a chop position angle of 65$^\circ$ (east of
north).}
\label{p5:obs_parameters}
\end{deluxetable}

\begin{deluxetable}{rrrrrrr}
\tablecolumns{7}
\tablewidth{0pc}
\tablecaption{Barnard 1 850 \micron\ Polarization Data}
\tablehead{
\colhead{$\Delta$ R.A.\tablenotemark{a}} & \colhead{$\Delta$ Dec.} & \colhead{$p$} & \colhead{$dp$} 
& \colhead{$\sigma_p$} & \colhead{$\theta$} & \colhead{$d\theta$} \\
\colhead{(\arcsec)} & \colhead{(\arcsec)} & \colhead{(\%)} & \colhead{(\%)} & \colhead{} & \colhead{($^\circ$)} & \colhead{($^\circ$)}}
\startdata
\multicolumn{7}{l}{\it Vectors associated with $I >$ 720 mJy beam$^{-1}$\tablenotemark{b}} \\
  $ -58.5$ & $-136.5$ &   3.74 &  0.56 &   6.7 & $ -85.3$ &  4.3 \\ 
  $ -64.5$ & $-136.5$ &   5.24 &  0.83 &   6.3 & $ -78.3$ &  4.5 \\ 
  $ -52.5$ & $-130.5$ &   1.86 &  0.50 &   3.7 & $  68.1$ &  7.7 \\ 
  $ -58.5$ & $-130.5$ &   2.36 &  0.46 &   5.2 & $  85.0$ &  5.6 \\ 
  $ -64.5$ & $-130.5$ &   3.85 &  0.59 &   6.5 & $ -80.9$ &  4.4 \\ 
  $ -70.5$ & $-130.5$ &   9.06 &  0.76 &  11.9 & $ -80.6$ &  2.4 \\ 
  $ -46.5$ & $-124.5$ &   2.49 &  0.63 &   3.9 & $  67.3$ &  7.3 \\ 
  $ -52.5$ & $-124.5$ &   3.10 &  0.46 &   6.7 & $  64.0$ &  4.3 \\ 
  $ -58.5$ & $-124.5$ &   1.79 &  0.44 &   4.1 & $  85.4$ &  7.1 \\ 
  $ -64.5$ & $-124.5$ &   2.29 &  0.61 &   3.7 & $ -72.8$ &  7.6 \\ 
  $ -46.5$ & $-118.5$ &   4.10 &  0.52 &   7.9 & $  79.3$ &  3.6 \\ 
  $ -52.5$ & $-118.5$ &   3.68 &  0.41 &   9.0 & $  79.8$ &  3.2 \\ 
  $ -58.5$ & $-118.5$ &   2.25 &  0.43 &   5.2 & $ -87.8$ &  5.5 \\ 
  $   1.5$ & $-112.5$ &   2.50 &  0.51 &   4.9 & $ -89.0$ &  5.9 \\ 
  $  -4.5$ & $-112.5$ &   2.47 &  0.46 &   5.4 & $ -88.7$ &  5.3 \\ 
  $ -10.5$ & $-112.5$ &   4.79 &  0.43 &  11.0 & $  66.7$ &  2.6 \\ 
  $ -16.5$ & $-112.5$ &   4.52 &  0.48 &   9.5 & $  89.1$ &  3.0 \\ 
  $ -22.5$ & $-112.5$ &   1.74 &  0.50 &   3.5 & $  85.6$ &  8.2 \\ 
  $ -46.5$ & $-112.5$ &   4.25 &  0.55 &   7.7 & $  84.6$ &  3.7 \\ 
  $ -52.5$ & $-112.5$ &   2.25 &  0.47 &   4.8 & $  78.3$ &  6.0 \\ 
  $ -58.5$ & $-112.5$ &   1.46 &  0.48 &   3.0 & $ -66.1$ &  9.5 \\ 
  $ -64.5$ & $-112.5$ &   2.13 &  0.54 &   3.9 & $ -80.1$ &  7.3 \\ 
  $  13.5$ & $-106.5$ &   4.00 &  0.50 &   8.0 & $  89.8$ &  3.6 \\ 
  $   7.5$ & $-106.5$ &   2.59 &  0.35 &   7.5 & $  89.2$ &  3.8 \\ 
  $   1.5$ & $-106.5$ &   1.05 &  0.32 &   3.3 & $  81.9$ &  8.7 \\ 
  $  -4.5$ & $-106.5$ &   1.57 &  0.32 &   4.8 & $ -82.1$ &  5.9 \\ 
  $ -10.5$ & $-106.5$ &   2.75 &  0.38 &   7.2 & $ -86.7$ &  4.0 \\ 
  $ -16.5$ & $-106.5$ &   4.26 &  0.44 &   9.7 & $ -86.8$ &  3.0 \\ 
  $ -22.5$ & $-106.5$ &   3.99 &  0.43 &   9.3 & $ -89.1$ &  3.1 \\ 
  $  13.5$ & $-100.5$ &   1.28 &  0.32 &   4.0 & $ -80.5$ &  7.2 \\ 
  $ -10.5$ & $-100.5$ &   2.08 &  0.33 &   6.4 & $  83.2$ &  4.5 \\ 
  $ -16.5$ & $-100.5$ &   2.86 &  0.41 &   7.0 & $ -78.2$ &  4.1 \\ 
  $   7.5$ & $ -94.5$ &   1.13 &  0.16 &   7.2 & $ -63.3$ &  4.0 \\ 
  $   1.5$ & $ -94.5$ &   1.01 &  0.15 &   6.6 & $ -49.0$ &  4.4 \\ 
  $ -10.5$ & $ -94.5$ &   1.28 &  0.36 &   3.5 & $  47.2$ &  8.1 \\ 
  $ -16.5$ & $ -94.5$ &   2.53 &  0.45 &   5.6 & $ -82.8$ &  5.1 \\ 
  $  19.5$ & $ -88.5$ &   2.38 &  0.41 &   5.8 & $  33.8$ &  4.9 \\ 
  $  -4.5$ & $ -88.5$ &   1.18 &  0.24 &   5.0 & $  46.4$ &  5.8 \\ 
  $ -10.5$ & $ -88.5$ &   3.40 &  0.41 &   8.4 & $  58.5$ &  3.4 \\ 
  $  13.5$ & $ -82.5$ &   1.09 &  0.28 &   3.9 & $  28.0$ &  7.4 \\ 
  $   7.5$ & $ -82.5$ &   1.15 &  0.18 &   6.5 & $  58.9$ &  4.4 \\ 
  $   1.5$ & $ -82.5$ &   1.35 &  0.18 &   7.4 & $  67.5$ &  3.9 \\ 
  $  -4.5$ & $ -82.5$ &   1.52 &  0.26 &   5.8 & $  68.0$ &  4.9 \\ 
  $ -10.5$ & $ -82.5$ &   4.02 &  0.46 &   8.8 & $  66.9$ &  3.3 \\ 
  $  13.5$ & $ -76.5$ &   1.53 &  0.37 &   4.1 & $ -50.2$ &  6.9 \\ 
  $   7.5$ & $ -76.5$ &   1.50 &  0.22 &   6.8 & $  73.7$ &  4.2 \\ 
  $   1.5$ & $ -76.5$ &   2.35 &  0.20 &  11.9 & $  66.6$ &  2.4 \\ 
  $  -4.5$ & $ -76.5$ &   1.91 &  0.27 &   7.0 & $  71.2$ &  4.1 \\ 
  $ -10.5$ & $ -76.5$ &   4.48 &  0.51 &   8.8 & $  65.5$ &  3.2 \\ 
  $  13.5$ & $ -70.5$ &   3.00 &  0.64 &   4.7 & $ -80.1$ &  6.1 \\ 
  $   7.5$ & $ -70.5$ &   4.50 &  0.37 &  12.1 & $ -87.4$ &  2.4 \\ 
  $   1.5$ & $ -70.5$ &   1.98 &  0.28 &   7.2 & $  76.8$ &  4.0 \\ 
  $  -4.5$ & $ -70.5$ &   1.35 &  0.34 &   3.9 & $  64.7$ &  7.3 \\ 
  $   7.5$ & $ -64.5$ &   5.30 &  0.65 &   8.2 & $  87.8$ &  3.5 \\ 
  $   1.5$ & $ -64.5$ &   4.13 &  0.45 &   9.3 & $  77.1$ &  3.1 \\ 
  $  -4.5$ & $ -64.5$ &   2.98 &  0.51 &   5.9 & $  71.7$ &  4.9 \\ 
  $ -28.5$ & $  13.5$ &   3.56 &  0.42 &   8.5 & $  61.2$ &  3.4 \\ 
  $ -34.5$ & $  13.5$ &   1.25 &  0.37 &   3.4 & $  72.4$ &  8.4 \\ 
  $ -22.5$ & $  19.5$ &   2.04 &  0.47 &   4.4 & $ -56.5$ &  6.5 \\ 
  $ -28.5$ & $  19.5$ &   1.22 &  0.33 &   3.7 & $   9.7$ &  7.8 \\ 
  $ -34.5$ & $  19.5$ &   1.41 &  0.25 &   5.6 & $  27.9$ &  5.1 \\ 
  $ -40.5$ & $  19.5$ &   1.45 &  0.22 &   6.5 & $  19.8$ &  4.4 \\ 
  $ -22.5$ & $  25.5$ &   3.12 &  0.43 &   7.2 & $ -24.0$ &  4.0 \\ 
  $ -28.5$ & $  25.5$ &   2.62 &  0.31 &   8.3 & $  24.8$ &  3.4 \\ 
  $ -34.5$ & $  25.5$ &   1.77 &  0.19 &   9.1 & $  31.6$ &  3.1 \\ 
  $ -40.5$ & $  25.5$ &   1.69 &  0.17 &   9.9 & $  31.9$ &  2.9 \\ 
  $ -28.5$ & $  31.5$ &   1.83 &  0.31 &   5.8 & $  30.0$ &  4.9 \\ 
  $ -34.5$ & $  31.5$ &   1.34 &  0.18 &   7.4 & $  44.9$ &  3.9 \\ 
  $ -40.5$ & $  31.5$ &   1.86 &  0.20 &   9.4 & $  47.0$ &  3.0 \\ 
  $ -46.5$ & $  31.5$ &   1.88 &  0.30 &   6.3 & $  50.3$ &  4.5 \\ 
  $ -52.5$ & $  31.5$ &   4.29 &  0.48 &   9.0 & $  64.6$ &  3.2 \\ 
  $ -34.5$ & $  37.5$ &   2.15 &  0.29 &   7.4 & $  47.1$ &  3.9 \\ 
  $ -40.5$ & $  37.5$ &   1.77 &  0.30 &   5.9 & $  46.9$ &  4.9 \\ 
  $ -46.5$ & $  37.5$ &   1.58 &  0.38 &   4.1 & $  40.6$ &  6.9 \\ 
\multicolumn{7}{l}{\it Vectors associated with $I <$ 720 mJy beam$^{-1}$\tablenotemark{b}} \\
  $ -10.5$ & $-139.5$ &   3.44 &  0.81 &   4.2 & $  47.8$ &  6.8 \\ 
  $ -22.5$ & $-139.5$ &   4.58 &  0.87 &   5.3 & $ -78.7$ &  5.5 \\ 
  $ -34.5$ & $-139.5$ &  11.66 &  0.93 &  12.6 & $  88.1$ &  2.3 \\ 
  $ -46.5$ & $-139.5$ &   6.88 &  0.52 &  13.2 & $ -70.8$ &  2.2 \\ 
  $ -58.5$ & $-139.5$ &   4.05 &  0.38 &  10.6 & $ -65.0$ &  2.7 \\ 
  $ -70.5$ & $-139.5$ &   8.00 &  0.64 &  12.4 & $ -73.1$ &  2.3 \\ 
  $  13.5$ & $-127.5$ &   8.54 &  0.96 &   8.9 & $  89.9$ &  3.2 \\ 
  $   1.5$ & $-127.5$ &   9.42 &  0.57 &  16.5 & $ -81.3$ &  1.7 \\ 
  $ -10.5$ & $-127.5$ &   3.99 &  0.41 &   9.7 & $  87.1$ &  3.0 \\ 
  $ -22.5$ & $-127.5$ &   3.64 &  0.44 &   8.2 & $  73.3$ &  3.5 \\ 
  $ -34.5$ & $-127.5$ &   8.27 &  0.45 &  18.5 & $ -86.2$ &  1.5 \\ 
  $ -46.5$ & $-127.5$ &   2.94 &  0.33 &   8.9 & $  67.2$ &  3.2 \\ 
  $ -82.5$ & $-127.5$ &   5.19 &  0.68 &   7.7 & $ -88.7$ &  3.7 \\ 
  $  25.5$ & $-115.5$ &   5.15 &  0.64 &   8.0 & $ -87.1$ &  3.6 \\ 
  $  13.5$ & $-115.5$ &   8.13 &  0.43 &  18.7 & $  82.0$ &  1.5 \\ 
  $   1.5$ & $-115.5$ &   2.91 &  0.30 &   9.6 & $ -85.2$ &  3.0 \\ 
  $ -22.5$ & $-115.5$ &   2.37 &  0.28 &   8.3 & $  84.0$ &  3.4 \\ 
  $ -34.5$ & $-115.5$ &   1.18 &  0.32 &   3.7 & $  56.5$ &  7.7 \\ 
  $ -82.5$ & $-115.5$ &   1.77 &  0.58 &   3.1 & $  86.0$ &  9.4 \\ 
  $  25.5$ & $-103.5$ &   6.52 &  0.46 &  14.1 & $  74.8$ &  2.0 \\ 
  $ -34.5$ & $-103.5$ &   4.63 &  0.31 &  15.1 & $  81.2$ &  1.9 \\ 
  $ -46.5$ & $-103.5$ &   2.64 &  0.30 &   8.7 & $ -85.3$ &  3.3 \\ 
  $ -70.5$ & $-103.5$ &   1.46 &  0.42 &   3.5 & $ -89.6$ &  8.2 \\ 
  $ -82.5$ & $-103.5$ &   2.14 &  0.63 &   3.4 & $  87.1$ &  8.4 \\ 
  $ -22.5$ & $ -91.5$ &   3.11 &  0.31 &  10.2 & $  89.5$ &  2.8 \\ 
  $ -34.5$ & $ -91.5$ &   3.41 &  0.36 &   9.4 & $ -87.3$ &  3.0 \\ 
  $ -46.5$ & $ -91.5$ &   1.16 &  0.34 &   3.4 & $  84.4$ &  8.3 \\ 
  $ -58.5$ & $ -91.5$ &   2.03 &  0.32 &   6.2 & $  79.1$ &  4.6 \\ 
  $ -70.5$ & $ -91.5$ &   1.65 &  0.46 &   3.6 & $  22.2$ &  7.9 \\ 
  $ -82.5$ & $ -91.5$ &   9.54 &  0.92 &  10.4 & $ -17.0$ &  2.7 \\ 
  $  25.5$ & $ -79.5$ &   3.43 &  0.62 &   5.5 & $  88.8$ &  5.2 \\ 
  $ -22.5$ & $ -79.5$ &   3.92 &  0.51 &   7.7 & $  75.4$ &  3.7 \\ 
  $ -34.5$ & $ -79.5$ &   3.78 &  0.46 &   8.3 & $ -85.8$ &  3.4 \\ 
  $ -46.5$ & $ -79.5$ &   4.09 &  0.36 &  11.5 & $  80.9$ &  2.5 \\ 
  $ -58.5$ & $ -79.5$ &   1.67 &  0.32 &   5.3 & $  35.8$ &  5.5 \\ 
  $ -70.5$ & $ -79.5$ &   3.06 &  0.43 &   7.2 & $  54.1$ &  4.0 \\ 
  $ -10.5$ & $ -67.5$ &   3.89 &  0.31 &  12.7 & $  83.2$ &  2.3 \\ 
  $ -22.5$ & $ -67.5$ &   6.48 &  0.62 &  10.5 & $  83.5$ &  2.7 \\ 
  $ -46.5$ & $ -67.5$ &   2.37 &  0.41 &   5.7 & $  82.5$ &  5.0 \\ 
  $ -58.5$ & $ -67.5$ &   4.19 &  0.33 &  12.8 & $  57.9$ &  2.2 \\ 
  $ -70.5$ & $ -67.5$ &   3.29 &  0.49 &   6.7 & $  15.0$ &  4.3 \\ 
  $  13.5$ & $ -55.5$ &   7.64 &  0.86 &   8.9 & $   0.1$ &  3.2 \\ 
  $   1.5$ & $ -55.5$ &   2.23 &  0.43 &   5.2 & $ -76.8$ &  5.5 \\ 
  $ -10.5$ & $ -55.5$ &   2.71 &  0.41 &   6.6 & $  70.5$ &  4.4 \\ 
  $ -22.5$ & $ -55.5$ &   8.82 &  0.72 &  12.2 & $ -85.7$ &  2.3 \\ 
  $ -46.5$ & $ -55.5$ &   3.04 &  0.98 &   3.1 & $  56.5$ &  9.2 \\ 
  $ -58.5$ & $ -55.5$ &   4.43 &  0.81 &   5.4 & $ -72.3$ &  5.3 \\ 
  $   1.5$ & $ -43.5$ &   2.54 &  0.72 &   3.5 & $  74.5$ &  8.1 \\ 
  $ -10.5$ & $ -43.5$ &   2.53 &  0.59 &   4.3 & $  86.9$ &  6.7 \\ 
  $ -22.5$ & $ -43.5$ &   7.08 &  0.90 &   7.8 & $  78.9$ &  3.7 \\ 
  $   1.5$ & $ -31.5$ &   6.44 &  0.70 &   9.2 & $ -70.8$ &  3.1 \\ 
  $ -10.5$ & $ -31.5$ &   5.43 &  0.59 &   9.1 & $ -67.9$ &  3.1 \\ 
  $ -22.5$ & $ -31.5$ &   9.13 &  0.94 &   9.7 & $  90.0$ &  3.0 \\ 
  $   1.5$ & $ -19.5$ &   8.57 &  0.79 &  10.9 & $ -84.0$ &  2.6 \\ 
  $ -10.5$ & $ -19.5$ &   4.18 &  0.68 &   6.1 & $ -60.5$ &  4.7 \\ 
  $ -22.5$ & $ -19.5$ &   4.86 &  0.79 &   6.1 & $  61.7$ &  4.7 \\ 
  $ -34.5$ & $ -19.5$ &   2.73 &  0.67 &   4.1 & $ -84.6$ &  7.0 \\ 
  $ -46.5$ & $ -19.5$ &   4.25 &  0.76 &   5.6 & $ -78.6$ &  5.1 \\ 
  $  13.5$ & $  -7.5$ &   3.89 &  0.78 &   5.0 & $ -68.9$ &  5.7 \\ 
  $   1.5$ & $  -7.5$ &   3.95 &  0.76 &   5.2 & $ -80.3$ &  5.5 \\ 
  $ -10.5$ & $  -7.5$ &   3.63 &  0.58 &   6.3 & $ -89.6$ &  4.6 \\ 
  $ -22.5$ & $  -7.5$ &   4.78 &  0.54 &   8.9 & $ -79.0$ &  3.2 \\ 
  $ -34.5$ & $  -7.5$ &   4.93 &  0.45 &  11.0 & $ -88.6$ &  2.6 \\ 
  $ -46.5$ & $  -7.5$ &   4.13 &  0.62 &   6.7 & $ -79.5$ &  4.3 \\ 
  $   1.5$ & $   4.5$ &   4.15 &  0.74 &   5.6 & $  66.9$ &  5.1 \\ 
  $ -10.5$ & $   4.5$ &   4.05 &  0.55 &   7.3 & $ -74.2$ &  3.9 \\ 
  $ -22.5$ & $   4.5$ &   1.25 &  0.36 &   3.5 & $ -51.8$ &  8.2 \\ 
  $ -34.5$ & $   4.5$ &   1.86 &  0.30 &   6.2 & $  54.0$ &  4.6 \\ 
  $ -58.5$ & $   4.5$ &   7.14 &  0.87 &   8.2 & $  42.9$ &  3.5 \\ 
  $  13.5$ & $  16.5$ &   8.55 &  0.93 &   9.2 & $ -58.2$ &  3.1 \\ 
  $   1.5$ & $  16.5$ &   3.23 &  0.65 &   5.0 & $ -63.8$ &  5.8 \\ 
  $ -10.5$ & $  16.5$ &   4.10 &  0.45 &   9.1 & $ -59.4$ &  3.2 \\ 
  $ -58.5$ & $  16.5$ &   4.97 &  0.63 &   7.9 & $ -86.8$ &  3.6 \\ 
  $   1.5$ & $  28.5$ &   7.91 &  0.73 &  10.8 & $ -84.6$ &  2.6 \\ 
  $ -10.5$ & $  28.5$ &   4.22 &  0.47 &   9.0 & $ -77.2$ &  3.2 \\ 
  $ -58.5$ & $  28.5$ &   3.48 &  0.43 &   8.0 & $  68.6$ &  3.6 \\ 
  $ -10.5$ & $  40.5$ &   5.84 &  0.76 &   7.7 & $ -61.6$ &  3.7 \\ 
  $ -22.5$ & $  40.5$ &   2.24 &  0.39 &   5.8 & $  -3.1$ &  5.0 \\ 
  $ -58.5$ & $  40.5$ &   4.96 &  0.54 &   9.1 & $  72.1$ &  3.1 \\ 
  $ -34.5$ & $  52.5$ &   2.08 &  0.69 &   3.0 & $ -71.1$ &  9.5 \\ 
  $ -58.5$ & $  52.5$ &   2.88 &  0.72 &   4.0 & $ -47.1$ &  7.2 \\ 
\enddata
\tablenotetext{a}{Positional offsets are given from the J2000.0
coordinates $\alpha = 03^{\rm h}33^{\rm m}$20\fs9 and $\delta =
+31$\degr09\arcmin03\farcs7 ($\alpha = 03^{\rm h}30^{\rm m}$15\fs0 and
$\delta = +30$\degr59\arcmin00\farcs0 in B1950.0).  Vectors are binned
to 12\arcsec\ sampling below the chosen threshold in total intensity
and 6\arcsec\ sampling above.  The total intensity at each vector
position exceeds 20\% of the faintest compact peak, B1-d.  This
minimizes the chances of systematic effects from chopping to a
reference position, as discussed in Appendix A of Paper II.}
\tablenotetext{b}{Using a calibration factor of 480 Jy beam$^{-1}$ V$^{-1}$.}
\label{p5:allthedata}
\end{deluxetable}

\begin{deluxetable}{lcccl}
\tablecolumns{5}
\tablewidth{0pc}
\tablecaption{Peak flux densities at 850 \micron}
\tablehead{
\colhead{Core} & \colhead{R.A.} & \colhead{Dec.} & \colhead{$S_{peak}$} & \colhead{Notes} \\
& \colhead{(J2000)} & \colhead{(J2000)} & \colhead{(\jybeam)} & }
\startdata
B1-a & $03^{\rm h}33^{\rm m}$16\fs4 & $+$31\degr07\arcmin51\arcsec\ & 0.68 & IRAS 03301+3057 (B1-IRS)\\
B1-bS & $03^{\rm h}33^{\rm m}$21\fs3 & $+$31\degr07\arcmin28\arcsec\ & 2.5 & \citet{hir99} \\
B1-c & $03^{\rm h}33^{\rm m}$17\fs7 & $+$31\degr09\arcmin31\arcsec\ & 3.0 & first continuum detection \\
B1-d & $03^{\rm h}33^{\rm m}$16\fs2 & $+$31\degr06\arcmin49\arcsec\ & 1.1 & first continuum detection \\
\enddata
\label{p5:B1peaks}
\end{deluxetable}

\begin{deluxetable}{cccrccc}
\tablecolumns{7}
\tablewidth{0pc}
\tablecaption{Magnetic Properties of Barnard 1}
\tablehead{
\colhead{$\sigma_{B_z}$\tablenotemark{a}} & \colhead{$i$\tablenotemark{b}} & \colhead{$B_0$} & \colhead{$N$} & \colhead{$\sigma_B$} & \colhead{$<B^2>^{-1/2}$} & \colhead{\bf{M$_n$/M$_u$}} \\
\colhead{(\mG)} & \colhead{(degrees)} & \colhead{($\mu$G)} & & \colhead{(\mG)} & \colhead{(\mG)} & }
\startdata
5.4 & 31 & 31 & 1 & 5.4 & 32.4 & 0.09 \\
    &    &    & 10 & 17.1 & 42.9 & 0.91 \\
10.8 & 50 & 42 & 1 & 10.8 & 46.0 & 0.20 \\
     &    &    & 10 & 34.2 & 72.6 & 2.00 \\
16.2 & 61 & 56 & 1 & 16.2 & 62.6 & 0.25 \\
     &    &    & 10 & 51.3 & 105 & 2.50 \\
21.6 & 68 & 71 & 1 & 21.6 & 80.3 & 0.28 \\
     &    &    & 10 & 68.4 & 138 & 2.80 \\
\enddata
\tablenotetext{a}{For all calculations, the following values were used: $|B_{0_z}|=27$ \mG; $s=0.33$ radians.}
\tablenotetext{b}{Inclination is measured from the line of sight.}
\label{p5:Bproperties}
\end{deluxetable}

\clearpage

\end{document}